\documentclass[runningheads]{llncs}
\usepackage[T1]{fontenc}
\usepackage{graphicx}
\usepackage{xcolor}
\usepackage{amsmath}
\usepackage{listings}
\usepackage[title]{appendix}

\newcommand{\OCLtext}{}

  {\begin{minipage}{#1} \OCLtext }%
  {\end{minipage}}
\newenvironment{ocl-boxed}[1][\linewidth]%
  {\begin{Sbox}\begin{minipage}{#1} \OCLtext }%
  {\end{minipage}\end{Sbox}{\setlength\fboxsep{7pt}\cornersize{0.3}\ovalbox{\TheSbox}}}

\begin{document}
\title{B-OCL: An Object Constraint Language Interpreter in Python}
%
 \author{Fitash UL HAQ\inst{1}\orcidID{0000-0003-2253-9085} \and
 Jordi Cabot\inst{1,2}\orcidID{0000-0003-2418-2489}}
 \authorrunning{F. UL HAQ et al.}
%
 \institute{Luxembourg Institute of Science and Technology, Luxembourg
 \email{fitash.ulhaq@list.lu}\\
 \and
 University of Luxembourg, Luxembourg\\
 \email{jordi.cabot@list.lu}}
\maketitle              
\begin{abstract}
The Object Constraint Language (OCL) has been widely used in the modeling community to complement software models for precisely defining constraints and business rules for the modeled systems.
There is a limited number of tools supporting the definition and interpretation of OCL constraints, even less for a  Python-based modelling approaches.

In this paper, we introduce an OCL interpreter for Python.
The interpreter has two components: parser and evaluator.
We implement the OCL metamodel as a set of Python classes and design the grammar for the parser using the state-of-the-art ANTLR parser generator.
The parser generates the syntax tree, that conforms with the OCL metamodel, after parsing each part of the OCL constraint.
The evaluator then interprets the constraints using this syntax tree and the object diagram. In the end, the interpreter reports the result for all the constraints.  

\end{abstract}

\section{Introduction}\label{sec:intro}
Object Constraint Language (OCL) is a formal language
widely used in the modelling community to complement the (UML) models with textual constraints that precisely define additional business rules for the system.
In practice, OCL is used to define different constructs on the software models, such as invariants, pre/post conditions, and constraints. 
There are several benefits for using OCL such as: (1) validation and verification of models, (2) better communication with stakeholders, (3) increased precision and formality.

\textbf{Motivation:}
Only a few open source and commercial tools are available for interpreting OCL constraints, such as Eclipse OCL~\cite{damus2002ocl}, USE~\cite{gogolla2007use}, MagicDraw~\cite{Dassault_Systemes_2024}, Oclarity~\footnote{http://www.empowertec.de/products/oclarity}  and IOCL~\cite{hammad2017iocl} and some of them are  abandoned. 
Moreover, most target Java-based modeling approaches.  Two small exceptions for Python would be 
pyalaocl\footnote{https://pyalaocl.readthedocs.io/en/latest/} provides functionality for writing  constraints in Python using an OCL-like syntax.  Similarly, the compiler pyecoreocl\footnote{https://github.com/aranega/pyecoreocl} allows OCL expressions to be written as part of  Python programs. Neither of the two provide an OCL interpreter.

This lack of OCL support in Python limits the creation and adoption of expressive model-based solutions for Python.  Such Python-based modeling solutions would be specially useful to integrate software modeling solutions with AI support, which is mostly Python-based as well.  



To solve this need, in this paper, we introduce a Python-based OCL interpreter to specify and evaluate OCL constraints as part of modeling projects created with BESSER~\cite{besser}.  
BESSER is an open-source state-of-the-art low-code development tool written in Python and it is designed to model, generate, personalize, and deploy smart and complex software systems. 
As a new component of this platform, our OCL interpreter is called B-OCL (short for BESSER OCL). 


The rest of the paper is as follows: 
Section~\ref{sec:nutshell} presents OCL in a nutshell.
Section~\ref{sec:overview} presents an overview of the B-OCL interpreter.
Section~\ref{sec:grammer} provides a brief description of Parser. 
Section~\ref{sec:evaluator} presents the description of Evaluator.
Section~\ref{sec:tool} presents the tool availability and limitations of current version of OCL interpreter.
Finally, section \ref{sec:conclusion} concludes the paper and discusses the future work.

\section{OCL in a nutshell}\label{sec:nutshell}
The Object Constraints Language (OCL) is a general-purpose declarative language designed to define constraints on models. It was introduced to overcome the limitations (i.e., specifying aspects of a system precisely) faced by visual models. OCL was first introduced by IBM in 1995, then it was adopted by the Object Management Group (OMG) in 1997, and it is now part of the UML standard. Initially, it was considered only for UML models, but gradually, it became an important and essential part of the Model-driven engineering (MDE) community.

OCL is used to define different types of expressions that include (1) Invariants, which are defined to specify all necessary conditions that must be satisfied in each possible instantiation of the model, (2) Initialization expressions, which are defined to initialize the class properties, (3) Derivation expressions, which are defined to specify how the elements must be computed, (4) Query expressions, which query the system and return the information to the user, and (5) last but the least, the Operation expressions (i.e., pre/post conditions) which must be satisfied before and after the operation respectively.
 
\begin{figure}
\centering
\includegraphics[scale=0.65
]{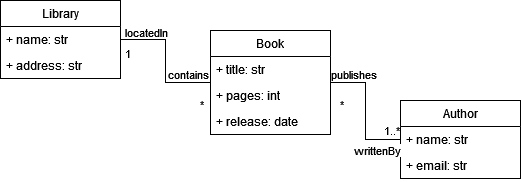}
\caption{Library Class Diagram} 
\label{fig:LIB-CD}
\end{figure}

For the purpose of exemplification throughout this paper, we use a simple example of a library class diagram as shown in figure~\ref{fig:LIB-CD}. This class diagram has three classes: (1) Library with attributes name and address, (2) Book with attributes title, pages, and release date, and (3) Author with attributes name and email. One simple constraint can be that the number of pages of the Book should be greater than 0. This constraint is shown in~\ref{eq:invBookConst}:

\begin{equation}\label{eq:invBookConst}
    \textbf{context}~ Book ~\textbf{inv}~ invBook: self.pages~ >~0
\end{equation}


OCL also supports defining conditional expressions on the classes, such as if-then-else expressions.  
Constraint\ref{OCL:IFthenElse} shows a if-then-else constraint which states that if the name of the library is `Children Library' then the number of pages for all instances of the book class should be less than 100, else True. 

\begin{equation}
\begin{split}
\textbf{context}~Library~\textbf{inv}~Constraint2: ~if~self.name = `Children~Library'~then
\\
~self.contains->forAll( i\_book~:~Book~|
    i\_book.pages <= 100 )   
    \\
    ~else~
    True
    ~endif
    \end{split}
    \label{OCL:IFthenElse}
\end{equation}

In addition, OCL also supports iteration-based expressions such as forAll, select, collect, reject, exists, etc, as also seen in the previous example where we go through all the library books to check they all satisfy the condition. 



\section{Overview}\label{sec:overview}

\begin{figure}[h]
\includegraphics[width=\linewidth]{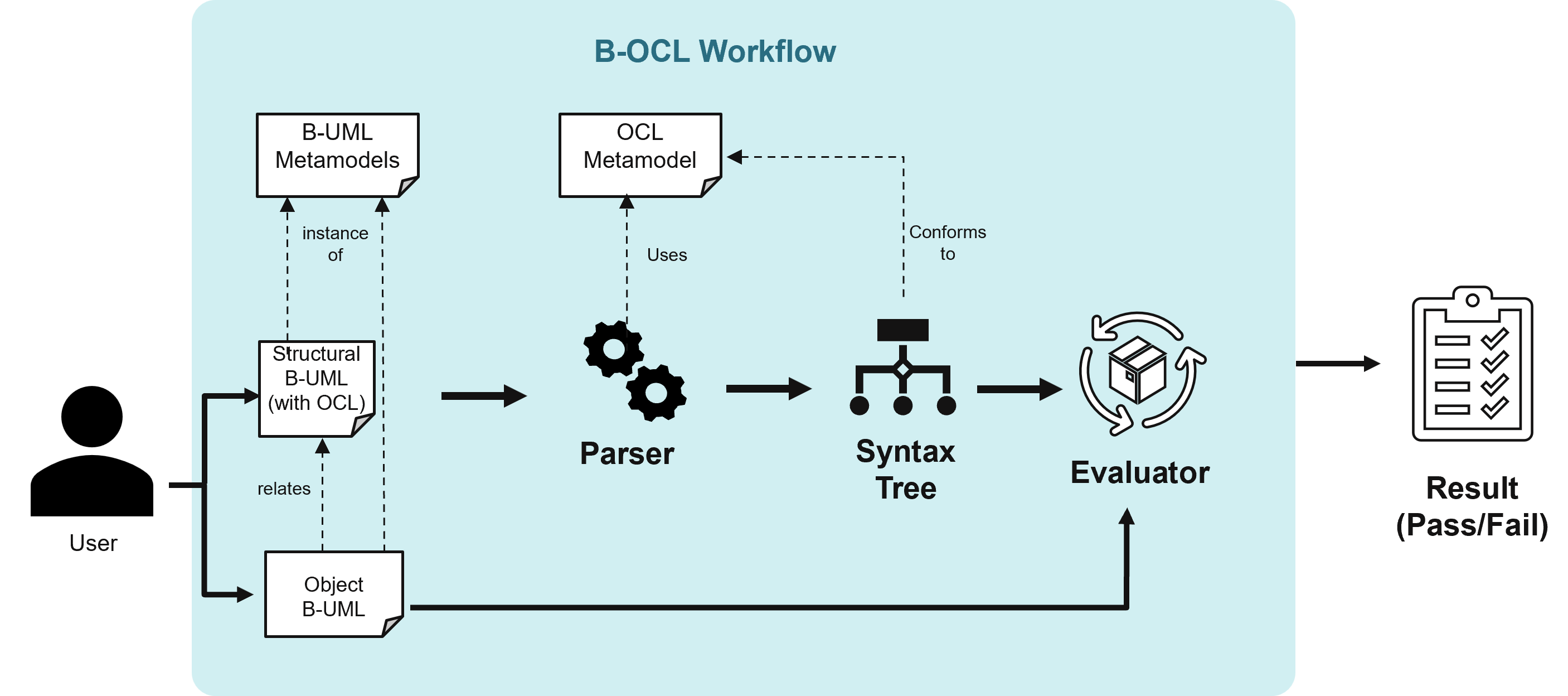}
\caption{Workflow of B-OCL interpreter} 
\label{fig:overview}
\end{figure}
Figure~\ref{fig:overview} shows the overview of the B-OCL interpreter.
The user provides the models in B-UML (BESSER Universal Modeling Language)~\cite{besser} format. B-UML is heavily inspired by UML\cite{specification2017omg}. Hence, B-UML models can basically be regarded as standard UML models
The user provides two B-UML models:
(1) a B-UML structural model that contains the domain model, together with a set of  OCL. 
(2) a B-UML object model that contains a relevant scenario the users would like to test expressed as a set of instantiated objects of the structural model




The interpreter consists of two main components: 
(1) \textbf{Parser}, that parses OCL constraints, iteratively, from the structural model and creates a syntax tree which conforms to OCL metamodel.
(2) \textbf{Evaluator}, that uses the syntax tree and the object model (i.e., an instance of the structural model) to evaluate the OCL constraint. 

Note that the object model is optional. The designer can just use the parser to check the syntactic correcness of the OCL constraints and create a combined UML+OCL syntax tree (conforming to the UML and OCL metamodels) that could be used as input for code-generation or other types of model manipulation operations. 
Next, we explain the parser and evaluator in section~\ref{sec:grammer} and section~\ref{sec:evaluator} respectively.



\section{Parser}\label{sec:grammer}
In order to create parser for OCL, we first define the OCL metamodel using the OCL specifications\footnote{https://www.omg.org/spec/OCL/2.4}. We also define the grammar for parsing the OCL constraints using the defined metamodel.
Listing \ref{lst:grammar} shows an excerpt of the grammar defined for OCL constraints. 
In this example, the oclFile is the entry point for all the constraints, which consists of two parts: (1) one \textit{contextDeclaration} that handles the context part of the OCL constraint, which specifies the class on which the constraint is applied and (2) \textit{expression} that is the condition part of the OCL constraint which should be true for all instances of the class.
Different types of expressions are supported by the grammar, including but not limited to binary expressions, if-else expressions, and loop expressions. More details on the types of expressions supported by the grammar can be found in the repository\footnote{"https://github.com/BESSER-PEARL/BESSER/tree/master/besser/BUML/notations/ocl"}.


\lstset{
    backgroundcolor=\color{yellow!10},%
    numbers=left, numberstyle=\tiny, stepnumber=2, numbersep=5pt,%
    }%

\lstdefinestyle{grammar}{
    basicstyle=\small\ttfamily\color{black},%
    breaklines=true,
    moredelim=[s][\color{green!50!black}\ttfamily]{'}{'},
    moredelim=*[s][\color{black}\ttfamily]{options}{\}},
    commentstyle={\color{gray}\itshape},
    morecomment=[l]{//},
    emph={%
        oclFile
        },emphstyle={\color{blue}\ttfamily},
    alsoletter={:,|,;},%
    morekeywords={:,|,;},
    keywordstyle={\color{black}},
}

\begin{lstlisting}[style=grammar, caption={Grammer for parsing OCL}, label={lst:grammar}]
oclFile: contextDeclaration (expression  )* ; // Context Declarations
contextDeclaration:
     CONTEXT ID (COLON type)? LBRACE? constraint* RBRACE? DoubleCOLON? functionCall? COLON? type?  LPAREN? ID? RPAREN? COLON? (DERIVE |BODY| Init | PRE | POST| Def)? COLON? expression? #ContextExp;

constraint: (INV | PRE | POST) ID? COLON expression SEMI? ;
functionCall: ID LPAREN (SingleQuote? expression SingleQuote? COMMA?)* RPAREN | ID LPAREN (ID COLON ID)* RPAREN
 | LPAREN(NUMBER COMMA?)* RPAREN;

 expression:
          (AND | OR )? binaryExpression expression? #binary
          | unaryExpression expression? #unary
          | IF expression  #ifExp
          | THEN expression #thenExp
          | ELSE expression #elseExp
          | ENDIF  expression? #endIfExp
          ...
// Other concepts omitted for brevity purposes ...
}
\end{lstlisting}

We use ANTLR~\cite{antlr}, a powerful tool widely used for language recognition, to generate a parser, lexer, and listener for OCL using the defined grammar.
The generated lexer first divides the OCL expression into tokens. The parser utilises these tokens and organizes them according to the grammar rules in tree format.  
The listener traverses the tree generated by the parser, identifying each node in the tree; the listener populates another tree using constructs from the OCL metamodel.   

\begin{figure}[h]
\includegraphics[width=\linewidth]{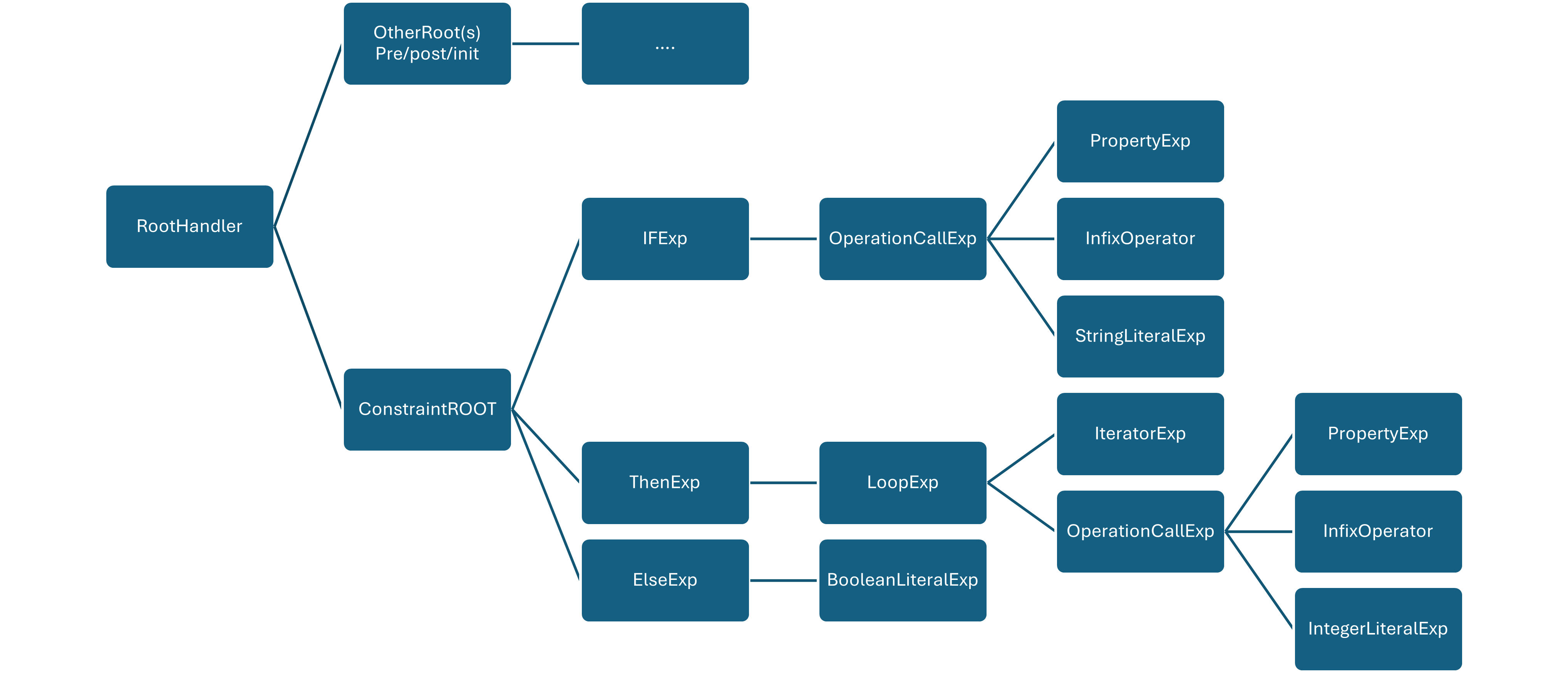}
\caption{Example of Tree generated by Parser
for the constraint shown in equation~\ref{OCL:IFthenElse}
} 
\label{fig:tree}
\end{figure}

Figure~\ref{fig:tree} shows a tree generated by the listener for a constraint with an if-then-else condition in the body (see  Constraint\ref{OCL:IFthenElse} in section~\ref{sec:nutshell}).
The listener starts by visiting the `if' part of the OCL constraint, then creating the IFExp from the OCL metamodel and adding it to the tree. In the next step, the listener visits the expression part of the IFExp, creates OperationCallExpression along with its parameters `PropertyExp,' `InfixOperator', and `StringLiteralExp', and adds it to the tree. The listener performs these steps repeatedly until all the constructs are visited. 
Similarly, the listener performs the same steps for `ThenExp' and `ElseExp' to populate the tree.

\section{Evaluator}\label{sec:evaluator}
The evaluator is the second component of the OCL interpreter. It uses the tree generated by the parser and object diagram provided by the user to determine whether the constraint is satisfied.
The evaluator, known as the B-OCL evaluator, is designed to support the majority of OCL constructs from the OCL metamodel required in the BESSER project. The interpreter is fully open source, so it can be easily extended to support other constructs.

The evaluator begins by parsing the tree and creating a logical expression depending on the type of node in the tree. This logical expression is populated using the values from the object diagram depending on the instances of the class to which the constraint is being applied. 

\begin{figure}
\centering
\includegraphics[scale=0.48]{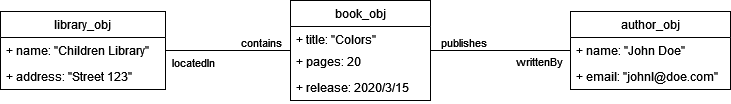}
\caption{Library Object Diagram} 
\label{fig:TP-OD}
\end{figure}

For purpose of exemplification, we define the object model of the library domain model explained in section~\ref{sec:nutshell}. The object model contains one library object, one book object and one author object as shown in figure~\ref{fig:TP-OD}. 
To explain the process, we interpret a simple constraint stating that the number of pages for every instance of book should be greater than 0 as shown in~\ref{eq:invBookConst}.
This constraint is transformed into the logical expression "\textit{self.pages} > 0," which is then populated with values from the object model, such as "20 > 0," which is then executed and results in the value `True'.

Depending upon the node in the tree generated by the parser, the
evaluator either updates the one primary logical expression or creates a sub-logical (or secondary logical) expression to evaluate before evaluating the primary expression. For example, in the case of IFExp, according to OCL specifications~\cite{OCL}, it is mandatory to have ThenExp and ElseExp, which results in two different paths from IFExp in the tree. In order to evaluate such an expression, the evaluator first creates a sub-logical expression using the expression part of IFExp and populates it with the values from the object diagram. 
This sub-logical expression is then evaluated to select the path of the tree. If the result of the sub-logical expression is `True', then the path of ThenExp is opted by the evaluator for evaluation; otherwise, the path of ElseExp is opted by the evaluator for evaluation.
Figure~\ref{fig:tree} shows the tree generated for IF-then-else expression shown in constraint~\ref{OCL:IFthenElse}.
Evaluator first creates a sub logical expression to verify if the name of library is `Children Library' using the values from object model; if it is True, the evaluator proceeds with updating the primary logical expression using the `then' part of the OCL constraint (i.e., following the ThenExp path of the generated tree) otherwise, it uses else part (i.e., following the ElseExp path of the generated tree) of the OCL constraint.

In the case of OCL constraints containing the iterator-based operations such as  `select', `forAll', `exists', and `reject',  the evaluator first solves the expression inside the collection operator and then updates the logical expression. 

The evaluator also supports simple binary expressions; binary expressions contain two operands separated by an operation. Operation can be equal to (`$=$'), not equal to(`$<>$'), greater than(`$>$'), less than(`$<$'), greater than and equal to(`$\geq$'), and less than and equal to(`$\leq$').

\section{Tool availability \& Limitations }\label{sec:tool}
B-OCL interpreter is available on Python Package Index (pypi) and can be installed using the following command.

\begin{lstlisting}[style=grammar, caption={Installation Command}, language = python ]
    pip install bocl
\end{lstlisting}

The interpreter is also available on github and can be installed from the source. 
To obtain the full code, including examples and tests, execute the following commands in terminal.
\begin{lstlisting}[style=grammar, caption={Building from source commands}, language = python ]
    git clone https://github.com/BESSER-PEARL/B-OCL-Interpreter
    cd B-OCL-Interpreter
    pip install -r requirements.txt
\end{lstlisting}
The interpreter utilizes the OCL metamodel and parser from the core BESSER~\cite{besser} repository. 

As mentioned in section~\ref{sec:intro}, the interpreter already supports most OCL constructs from OCL metamodel. 
However, some constructs are still not supported and are planned to be released in the future. These include initialization expressions, derived expressions, query operations, enumerations, collectionType, pre/post conditions, and messageState.
The rationale behind not supporting derived expressions and query operations is that such expressions can be reformulated into simpler expressions without affecting the outcome of the expression. Another limitation is handling of null value, which causes the three-valued logic problem~\cite{brucker2010extending,cabot2012object}.  As this is a complex problem with different interpretations within the OCL community, it is for now left as future work.  

\section{Conclusion and Future Work}\label{sec:conclusion}

In this paper, we present a state-of-the-art OCL interpreter specifically tailored for Python-based modelling approaches. 
The interpreter takes the domain model, containing the OCL constraints, and the object model  as input. The interpreter returns a list of all OCL constraints and their results on the object model, indicating whether they passed or failed.

As part of future work, we intend to extend the list of constructs supported for OCL. 
Furthermore, we plan to add formal verification support to the tool via an intermediate mapping from OCL to Z3~\cite{de2008z3} to check the correctness properties of the modeled OCL constraints, such as SAT.
In addition, we intend to bridge the gap with the semantic web community by transforming OCL expressions to Shapes Constraint Language (SHACL)~\cite{knublauch2015shapes}.
We also intend to add code generators: (1) to convert OCL constraints into SQL or other techs to maximize the ROI of writing OCL expressions, and (2) to generate code that enforces the constraints efficiently, checking only those constraints affected by the change in the system~\cite{cabot2009incremental}.


%
%
%
\bibliographystyle{splncs04}
\bibliography{references}

\begin{thebibliography}{10}
\providecommand{\url}[1]{\texttt{#1}}
\providecommand{\urlprefix}{URL }
\providecommand{\doi}[1]{https://doi.org/#1}

\bibitem{OCL}
\url{https://www.omg.org/spec/OCL/2.4/PDF}

\bibitem{Dassault_Systemes_2024}
 (Jun 2024), \url{https://www.3ds.com/products/catia/no-magic/magicdraw}

\bibitem{besser}
Alfonso, I., Conrardy, A., Sulejmani, A., Nirumand, A., Ul~Haq, F., Gomez-Vazquez, M., Sottet, J.S., Cabot, J.: Building besser: an open-source low-code platform. In: International Conference on Business Process Modeling, Development and Support. pp. 203--212. Springer (2024)

\bibitem{brucker2010extending}
Brucker, A.D., Krieger, M.P., Wolff, B.: Extending ocl with null-references: towards a formal semantics for ocl 2.1. In: Models in Software Engineering: Workshops and Symposia at MODELS 2009, Denver, CO, USA, October 4-9, 2009, Reports and Revised Selected Papers 12. pp. 261--275. Springer (2010)

\bibitem{cabot2012object}
Cabot, J., Gogolla, M.: Object constraint language (ocl): a definitive guide. In: International school on formal methods for the design of computer, communication and software systems, pp. 58--90. Springer (2012)

\bibitem{cabot2009incremental}
Cabot, J., Teniente, E.: Incremental integrity checking of uml/ocl conceptual schemas. Journal of Systems and Software  \textbf{82}(9),  1459--1478 (2009)

\bibitem{damus2002ocl}
Damus, C., S{\'a}nchez-Barbudo, A.: Ocl documentation (2002)

\bibitem{de2008z3}
De~Moura, L., Bj{\o}rner, N.: Z3: An efficient smt solver. In: International conference on Tools and Algorithms for the Construction and Analysis of Systems. pp. 337--340. Springer (2008)

\bibitem{gogolla2007use}
Gogolla, M., B{\"u}ttner, F., Richters, M.: Use: A uml-based specification environment for validating uml and ocl. Science of Computer Programming  \textbf{69}(1-3),  27--34 (2007)

\bibitem{specification2017omg}
Group, O.M.: {OMG} unified modeling language tm (omg {UML}) pp. 1--754 (2017)

\bibitem{hammad2017iocl}
Hammad, M., Yue, T., Wang, S., Ali, S., Nyg{\aa}rd, J.F.: Iocl: An interactive tool for specifying, validating and evaluating ocl constraints. Science of Computer Programming  \textbf{149}, ~3--8 (2017)

\bibitem{knublauch2015shapes}
Knublauch, H., Kontokostas, D.: Shapes constraint language (shacl). w3c editor’s draft. World Wide Web Consortium. http://w3c. github. io/data-shapes/shacl  (2015)

\bibitem{antlr}
Parr, T.: The definitive antlr 4 reference. The Definitive ANTLR 4 Reference pp. 1--326 (2013)

\end{thebibliography}
\newpage
\begin{subappendices}
\renewcommand{\thesection}{\Alph{section}}%

\section{Demonstration}

The core BESSER platform contains the OCL metamodel and parser, so it is possible to  annotate the B-UML models with OCL constraints. 
The B-OCL (BESSER Object Constraint Language) interpreter adds to that the capability to actually execute those constraints and evaluate them on top of an object model, where each object is an instance of a B-UML class. 
In order to define the OCL constraints, we need to define a domain model.

\subsection{First Step: Defining classes and associations in a Domain Model}

In the first step, we need a domain model. For the purpose of exemplification, we consider the same library domain model defined in section~\ref{sec:nutshell}. To revisit, the domain model has three classes: (1) Library, (2) Book, and (3) Author as shown in Figure~\ref{fig:LIB-CD}. 


The corresponding B-UML code for defining the classes and association in the diagram is shown in listing~\ref{lst:gen-code}.  Note that a designer is not expected to write this definition by hand. This Python script would be  generated from a textual or graphical UML representation of the Library model from Figure~\ref{fig:LIB-CD}. 

\begin{lstlisting}[style=grammar, caption={Code for Class Diagram}, label={lst:gen-code}]
t_int: PrimitiveDataType = PrimitiveDataType("int")
t_str: PrimitiveDataType = PrimitiveDataType("str")
t_date: PrimitiveDataType = PrimitiveDataType("date")

# Library attributes definition
library_name: Property = Property(name="name", type=t_str)
address: Property = Property(name="address", type=t_str)
# Library class definition
library: Class = Class (name="Library", attributes={library_name, address})

# Book attributes definition
title: Property = Property(name="title", type=t_str)
pages: Property = Property(name="pages", type=t_int)
release: Property = Property(name="release", type=t_date)
# Book class definition
book: Class = Class (name="Book", attributes={title, pages, release})

# Author attributes definition
author_name: Property = Property(name="name", type=t_str)
email: Property = Property(name="email", type=t_str)
# Author class definition
author: Class = Class (name="Author", attributes={author_name, email})

# Library-Book association definition
located_in: Property = Property(name="locatedIn",type=library, multiplicity=Multiplicity(1, 1))
contains: Property = Property(name="contains", type=book, multiplicity=Multiplicity(0, "*"))
lib_book_association: BinaryAssociation = BinaryAssociation(name="lib_book_assoc", ends={located_in, contains})

# Book-Author association definition
publishes: Property = Property(name="publishes", type=book, multiplicity=Multiplicity(0, "*"))
writed_by: Property = Property(name="writedBy", type=author, multiplicity=Multiplicity(1, "*"))
book_author_association: BinaryAssociation = BinaryAssociation(name="book_author_assoc", ends={writed_by, publishes})


\end{lstlisting}

\subsubsection{Defining OCL Constraints on the Domain Model}

After defining the classes and associations, the next step in definition of OCL constraints. For the purpose of exemplification, we have defined two constraints: (1) number of pages for each book should be greater than 0 in equation~\ref{ocl:playerAge}, and (2) library should contain atleast one book with number of pages less than 110 in  equation~\ref{ocl:centerPlay}.

\begin{equation}
\begin{split}
\textbf{context}~Book ~\textbf{inv} ~pageNumberInv: ~self.pages>~0
\end{split}
\label{ocl:playerAge}
\end{equation}

\begin{equation}
\begin{split}
\textbf{context}~Library~\textbf{inv}~ atLeastOneSmallBook:
self.contains~\\->~
select(i\_book~:~ Book ~|~ i\_book.pages~ <= ~110)
~
           \\->~size()~>~0
\end{split}
\label{ocl:centerPlay}
\end{equation}

And the corresponding B-UML code for OCL constraint is shown in Listing~\ref{lst:oclConsts}.

\begin{lstlisting}[style=grammar, caption={BUML Code for OCL constraints}, label={lst:oclConsts}]
    constraintBookPageNumber: Constraint = Constraint(name = "BookPageNumber",context=book,expression="context Book inv pageNumberInv: self.pages>0",language="OCL")

    constraintAtLeastOneSmallBook: Constraint = Constraint(name = "LibaryCollect",context=library,
    expression="context Library inv atLeastOneSmallBook: self.contains->select(i_book : Book | i_book.pages <= 110)->size()>0",language="OCL")
\end{lstlisting}

In the final step defining the domain model, we  combine all the classes, associations and constraints in to one domain model as shown in listing~\ref{lst:dm}. This completes the declaration of domain model in B-UML langauge.
\begin{lstlisting}[style=grammar, caption={Domain Model Declaration}, label={lst:dm}]
    library_model : DomainModel = DomainModel(name="Library model", types={library, book, author}, 
    associations={lib_book_association, book_author_association},
    constraints={constraintBookPageNumber, constraintAtLeastOneSmallBook})  
\end{lstlisting}

\subsection{Second Step: Defining an object Model}

In the second step, we define an object model for these constraints to be validated on. Using the same example of library class diagram, figure~\ref{fig:TP-OD} (in section~\ref{sec:evaluator}) shows the object diagram containing three objects: one library object, one book object and one author object.

\begin{lstlisting}[style=grammar, caption={Code for Object Diagram}, label={lst:gen-code-OD}]
# Library object attributes

library_obj_name: AttributeLink = AttributeLink(attribute=library_name, value=DataValue(classifier=t_str, value="Children Library"))
library_obj_address: AttributeLink = AttributeLink(attribute=address, value=DataValue(classifier=t_str, value="Street 123"))
# Library object
library_obj: Object = Object(name="library_obj", classifier=library, slots=[library_obj_name, library_obj_address])

# Book object attributes
book_obj_name: AttributeLink = AttributeLink(attribute=title, value=DataValue(classifier=t_str, value="Colors"))
book_obj_pages: AttributeLink = AttributeLink(attribute=pages, value=DataValue(classifier=t_int, value=20))
book_obj_release: AttributeLink = AttributeLink(attribute=release, value=DataValue(classifier=t_date, value=datetime.datetime(2020, 3, 15)))
# Book object
book_obj: Object = Object(name="book_obj", classifier=book, slots=[book_obj_name, book_obj_pages])


author_obj_name: AttributeLink = AttributeLink(attribute=author_name, value=DataValue(classifier=t_str, value="John Doe"))
author_obj_email: AttributeLink = AttributeLink(attribute=email, value=DataValue(classifier=t_str, value="john@doe.com"))
# Author object
author_obj: Object = Object(name="Author Object", classifier=author, slots=[author_obj_name, author_obj_email])


# Book object and Author object link
book_link_end1: LinkEnd = LinkEnd(name="book_end1", association_end=publishes, object=book_obj)
author_link_end: LinkEnd = LinkEnd(name="author_end", association_end=writed_by, object=author_obj)
author_book_link: Link = Link(name="author_book_link", association=book_author_association, connections=[book_link_end1,author_link_end])

# Book Library and Book object link
book_link_end2: LinkEnd = LinkEnd(name="book_end2", association_end=has, object=book_obj)
library_link_end: LinkEnd = LinkEnd(name="library_end", association_end=located_in, object=library_obj)
library_book_link: Link = Link(name="library_book_link", association=book_author_association, connections=[book_link_end2,library_link_end])

# Object model definition
object_model: ObjectModel = ObjectModel(name="Object model", instances={library_obj, author_obj, book_obj}, links={author_book_link, library_book_link})


\end{lstlisting}

Listing~\ref{lst:gen-code-OD} shows the B-UML code for object diagram.

\subsection{Third Step: Evaluating the constraints}

In the third and final step, the OCL constraints can be evaluated using the script shown in listing~\ref{lst:exe}.

\begin{lstlisting}[style=grammar, caption={Script for Executing the Evaluator}, language = python ,label={lst:exe}]
from models.library_object import library_model,object_model
from bocl.OCLWrapper import OCLWrapper

if __name__ == "__main__":
    wrapper = OCLWrapper(library_model, object_model)
    for constraint in library_model.constraints:
        print("Invariant:" + str(constraint.expression), end=":")
        res = None
        try:
            res = wrapper.evaluate(constraint)
            print('\x1b[0;30;35m' + str(res) + '\x1b[0m')

        except Exception as error:
            print('\x1b[0;30;41m' + 'Exception Occured! Info:' + str(error) + '\x1b[0m')
            res = None
        
    
\end{lstlisting}

The output after executing this script using python 3.9+ is shown in listing~\ref{lst:res}

\begin{lstlisting}[style=grammar, caption={Script for Executing the Evaluator} ,label={lst:res}]
Invariant:context Book inv pageNumberInv: self.pages>0:True
Invariant:context Library inv atLeastOneSmallBook:
self.contains->select(i_book: Book | i_book.pages <= 110)->size()>0:True
\end{lstlisting}

Using these three steps,  user can easily define and interpret OCL constraints on the B-UML models.

\end{subappendices}

\end{document}